\documentstyle[pre,aps,epsf,psfig,manuscript]{revtex}
\newcommand \be{\begin{equation}}
\newcommand \ee{\end{equation}}
\newcommand \ba{\begin{eqnarray}}
\newcommand \ea{\end{eqnarray}}
\newcommand \ep {\varepsilon}

\newcommand{\DK}[1]{\mbox{\boldmath$#1$}}
\newcommand{\BE}{\begin{equation}\label}
\newcommand{\BEQ}{\begin{eqnarray}\label}
\newcommand{\EE}{\end{equation}}
\newcommand{\EEQ}{\end{eqnarray}}

\begin{document}
\draft
\title{Electric microfield distribution in two-component plasmas. Theory and Simulations}

\author{J.~Ortner $^{a)}$\footnote{Corresponding author, Tel.: (+4930) 2093 7636, email: jens@physik.hu-berlin.de}, I. Valuev $^{b)}$, and W.~Ebeling$^{a)}$}
\address{$^{a)}${\it Institut f\"ur Physik, Humboldt Universit\"{a}t zu Berlin,\\ Invalidenstr. 110, D-10115 Berlin, Germany}\\$^{b)}${\it {Department of Molecular and Chemical Physics, Moscow Institute of Physics and Technology,\\ 141700 Dolgoprudny, Russia}}}
\date{to be published in Contr. Plasma Phys.}
\maketitle

\begin{abstract}
The distribution of the electric microfield at a charged particle moving in a two-component plasma is calculated. 
 The theoretical approximations are obtained via the parameter integration technique and using the screened pair approximation for the generalized radial distribution function. It is shown that the two-component plasma microfield distribution shows a larger probability of high microfield values than the corresponding distribution of the commonly used OCP model. The theory is checked by quasiclassical molecular-dynamics simulations. For the simulations a corrected Kelbg pseudopotential has been used.

\vspace*{0.5cm}

\vspace*{0.5cm}
\end{abstract}

\section{Introduction}

The purpose of this paper is the investigation of the microfield distribution in a two-component plasma at the position of a charged particle. 

The determination of the distribution of the electric microfield component created by one of the subsystems separately - electron or ion - is a well studied problem (for a review see \cite{dufty}). Holtsmark \cite{Holtsmark} reduced the line shape problem to the determination of the probability distribution of perturbing ionic electric microfield. In recent papers it was argued that the electric microfield low frequency part (due to the ion dynamics) also influences the fusion rates \cite{RomEb98} and the rates for the three-body electron-ion recombination \cite{Rom98} in dense plasmas. 
Holtsmark's work on the electric microfield distribution was restricted to ideal plasmas. The opposite limiting case of infinite coupling strength was considered by Mayer \cite{Mayer,Broyles} within the ion sphere model. Within this model the central ion undergoes harmonic oscillations around the center of the negatively charged ion sphere. This results in a Gaussian approximation for electric microfields at the ion position.
The nonideality of plasmas leads to quantitative
corrections to Holtsmark's result as shown by Baranger and Mozer
\cite{baranger} and Iglesias \cite{Iglesias} for the case of weakly coupled plasmas and  by Iglesias et al. \cite{ILM83} for the case of strongly coupled plasmas. In these papers it is shown that with increasing coupling strength $\Gamma$ the long tailed Holtsmark distribution is changed into the fast decaying Gaussian approximation. Here the coupling parameter $\Gamma= e^2/kTd$ is defined via the electron density $n_e$ ($ d = [3/4\pi n_e]^{1/3} $ is the average distance of the electrons).

In the cited papers the electric microfield created by one of the subsystems has been studied by an almost total neglect of the influence of the other subsystem. A common assumption is that the distribution of the high-frequency component (due to the electron dynamics) is the same as that of an electron gas with uniform neutralizing background. This is the so called model of the one component plasma (OCP). For the ion subsystem, in a first approximation, the electrons are assumed to move free through the plasma. Since the electron motion is much more rapid than the ion one, the electrons are treated as a smeared negative charged background. For simplicity this background charge was assumed to be uniform in the density and not to be distorted by the ion motion. This again is the OCP model.

A more realistic model should also take into account the variation of the background charge density. A background charge distribution which differs from a uniform distribution results in a screening of the ion motion, the screening strength is generally frequency dependent, e.g. it depends on the ion velocity. In a first approximation one might neglect the frequency dependence of the screening. Then one arrives at the model of an OCP on a polarizable background (POCP). In the theory of microfields this slightly more involved model is used to describe the low frequency part \cite{baranger,ecker}. However, both the OCP and the POCP fail to describe the correlations between the electron and the ion subsystem.  

To include the electron-ion correlations one has to consider the model of a two-component plasma (TCP). This paper is adressed to the electric microfield studies in an equilibrium two-component plasma. To our knowledge the electric microfield in a TCP has been studied only by Yan and Ichimaru \cite{YI86}. However, due to a couple of flaws contained in the paper of Yan and Ichimaru a further investigation is required. For simplicity we will restrict ourselves to the case of a two-component 
plasma which is anti-symmetric with respect to 
the charges ($e_- = - e_+$) and therefore symmetrical with respect to the densities
($n_+ = n_i = n_- = n_e$). Further, the theoretical investigations are carried out for arbitrary electron ion mass ratios. To simplify the numeric investigations we simulated so far only a mass symmetrical (nonrelativistic) electron-positron plasma with $m= m_+ = m_e$.   We study this - so far unrealistic - case 
of mass - symmetrical plasmas in order to save computer time in particle simulations. The mass-symmetrical model is well suited to check the quality of various analytical approximations. In addition, the results of the simulation are also applicable to the case of an electron-hole plasma in semiconductors.

As for the case of the OCP the  microfield distribution of a TCP in the weak coupling regime is approximated by the Holtsmark distribution.  However, coupled plasmas are important objects
in nature, laboratory experiments, and in 
technology \cite{gruenes-buch,ichimaru92,ksruegen}.

Therefore we are interested in the modification of the microfield distribution caused by the coupling of plasma particles. Both  theoretical investigations and semiclassical simulations are performed to study the microfield distribution in two-component plasmas.

In this paper the free charges (electron and ions) are simulated by a semiclassical dynamics based on effective potentials.  
The idea of the semiclassical method exploited in the numerical part of this paper is to incorporate
quantum-mechanical effects (in particular the
Heisenberg and the Pauli principle) by appropriate potentials. This method was pioneered by Kelbg, Dunn and Broyles, Deutsch and others \cite{Kelbg,DB67,Deutsch}. Several investigations were
devoted to the simulation of equilibrium two-component plasmas
\cite{norman,hansen2,PC94,Klakow,penman95}.
Being interested in semiclassical methods we mention explicitely the semiclassical 
simulations of two-component plasmas performed by Norman and by Hansen 
\cite{norman,hansen2}.

Certainly, such a semiclassical approach has several limits. For example, bound states cannot be described classically, therefore our methods are 
restricted to the subsystem of the free charges. However, this is not a very serious
restriction since most of the plasma properties are determined by the subsystem
of the free charges.

The semiclassical approach may be very usefull  to calculate a standard
macroscopic property such as the microfield distribution since it has a well defined classical limit. The advantage of such an approach is the relative simplicity of the algorithm.

\section{microfield distribution} \label{II}

Consider a two-component plasma consisting of $N_i$ ions of one species and $N_e=N_i$ electrons with masses $m_e$ and $m_i$ and charges $e_e=-e_i=-e$.  The total number of particles is $N=N_e+N_i$.  The plasma system with the total volume $V$ and temperature in energy units $T=1/\beta$  is described by the Hamilton operator

\be \label{H}
\hat{H}~=~\hat{K}~+~\hat{V}~=~\sum_{a=e,i} \sum_{i=1}^{N_a} \frac{\hat{\DK{p}}_{a,i}^2}{2 m_a} ~+~ \frac{1}{2} \sum_{a,b=e,i} \sum_{i=1}^{N_a} \sum_{j=1}^{N_b} \hat{v}_{ab}(\DK{r}_{a,i},\DK{r}_{b,j}) \quad.
\ee 

The interaction potential between two particles is given by the Coulomb potential

\be 
 v_{ab}(\DK{r_i},\DK{r_j})~=~\frac{e_a e_b}{|\DK{r_i} -\DK{r_j}|} \quad. \nonumber
\ee

The operator of electric field acting on a certain particle (hereafter called the first particle) is defined by the sum of single particle Coulomb field operators,

\BE{field}
\DK{E}=\sum_{j=2}^N \DK{E}_j(\DK{r}_{1j}) \quad,~~
\DK{E}_j(\DK{r}_{1j})=- \frac{e_j}{r_{1j}^3}\DK{r}_{1j}, ~~ r_{ij}=|\DK{r}_i-\DK{r}_j| \quad .
\ee

Define now the electric microfield distribution $W(\ep)$ as the probability of measuring an electric field  $\DK{\ep}$ equal to $\DK{E}$ at the probe charge position $\DK{r}_1$,
\be \label{WE}
W(\DK{\ep})=<\delta(\DK{\ep} - \DK{E})> \quad,
\ee
where $< \hat{A} >\,\,=\,\,(1/Z)\,{\rm Sp} \left( \hat{A} \exp[- \beta \hat{H}]\right)$ denotes the quantumstatistical average of the operator $\hat{A}$, and $Z={\rm Sp}  \exp[- \beta \hat{H}]$ is the partition sum of the plasma system.

We assume that our system is isotropic. Then we may rewrite  Eq.(\ref{WE}) as \cite{Broyles}
\be \label{PE}
P(\ep)=2 \pi^{-1} \ep \int _0^{\infty} \,dl\,\,l\,T(l)\,{\rm sin}(\ep l) \quad,
\ee
where $P(\ep)$ is related to $W(\ep)$ by $4 \pi W(\ep) \ep^2 d \ep = P(\ep) d \ep$,
and
\be \label{Fourier}
T(\DK{k})=<e^{i \DK{k} \cdot \DK{\ep}}>
\ee
is the Fourier transform of the microfield distribution function $W(\ep)$.

It is convenient to introduce the dimensionless quantity
\BE{dimless1}
\DK{F}=\frac{\DK{E}}{E_0} \quad ,
\EE
where $E_0$ is defined through the total density $n=N/V$ by
$E_0=2\pi (4/15)^{2/3}\,e\,n^{2/3}$. 
The probability distribution for the dimensionless field $F$ then becomes with $L=l E_0$,
\be \label{PF}
P(F)=2 \pi^{-1} F \int _0^{\infty} \,dL\,\,L\,T(L)\,{\rm sin}(F L) \quad.
\ee

Consider now some known limiting cases for the microfield distribution. In the weak coupling regime for $\Gamma \ll 1$ the Holtsmark distribution is applicable for the microfield distribution and we have
\BE{Holts}
T(L,\Gamma \ll 1) = T_H(L) = \exp[-L^{3/2}] \quad .
\ee

The other limiting case of strong coupling $\Gamma \gg 1$ is known so far only for the one-component plasma model. For the OCP the ion sphere model holds in the strong coupling regime. Within this model the charge will be attracted towards the center of its oppositely charged sphere of radius $ d = [3/4\pi n_e]^{1/3} $ and with average density $n_e$. The harmonic potential for the displacement of the center leads then to a Gaussian approximation for the distribution of the normalized electric field $F=E/E_{0,{\rm OCP}}$ at the charge ,
\be \label{Gauss}
P(F)=(2/\pi)^{1/2}(F^2/\tau^{3/2}) \,\, \exp(-F^2/2\tau) \quad,
\ee
where
\be \label{Gauss2}
\tau\,\,=\,\,\left( \,b \,\Gamma \, \right)^{-1} \; {x} {\rm coth} \, x~,~~x= (T \hbar^2/4 m_e e^4) \Gamma^3 \,\,,~b=\frac{4}{5}\left( \frac{2 \pi^2}{5} \right)^{1/3} \,\,.
\ee
The normalizing field strength for the OCP case should be expressed in terms of the electron density $n_e$ only, $E_{0,{\rm OCP}}=2\pi (4/15)^{2/3}\,e\,n_e^{2/3}$.
In the case of a classical one-component plasma $\hbar \to 0$ the parameter $\tau$ playing the role of an effective temperature in the Gaussian distribution Eq.(\ref{Gauss}) simplifies and reads
$\tau_{{\rm cl}}\,\,=\,\,{1}/{(b \Gamma)}$,
and Eq.(\ref{Gauss}) turns into the expression developed by Mayer \cite{Mayer}.

However, there is no commonly accepted generalization of the ion sphere model for the two-component plasma with charges of different signs. Moreover we will show that in the case of TCP there is not any analogue for the Gaussian distribution in strongly coupled OCP.

First we mention that the Fourier transform of the Gaussian distribution for electric microfield applicable in the strong coupling OCP regime equals
\BE{GaussF}
T(L,\Gamma \gg 1) = T_G(L) = \exp\left[- \frac{L^2 \tau}{2} \right] \quad .
\EE
Notice that the Taylor expansion of the Gaussian function $T_G(L)$ starts with
\BE{GaussTay}
T_G(L) = 1 \,\, - \,\, \frac{L^2 \tau}{2} \quad.
\EE

On the other hand it is possible to perform exact calculations for the leading terms in the small $L$ expansion of the Fourier transform $T(L)$. From the definition of $T(k)$ Eq. (\ref{Fourier}) it follows that
\BE{Taylor}
T(\DK{k} \to 0)~=~1~-~k^2\,\frac{<\ep^2>}{6}~+~k^4\,\frac{<\ep^4>}{120}~\pm \ldots \quad.
\EE
In Refs. \cite{ILM83} it was argued that it is necessary to incorporate the knowledge of the second moment $<\ep^2>$ into the calculation of microfield distributions in OCP. One might now try to generalize this idea to the case of a TCP. However, as can be easily seen the coefficient in the $k^2$ term $<\ep^2>$ diverges in the case of a TCP:
\BE{div}
<\ep^2>~=~<\, \sum_{j=2}^N \left(\,\DK{E}_j(\DK{r}_{1j})\,\right)^2 \;+ \sum_{j \neq k} \DK{E}_j(\DK{r}_{1j})\cdot \DK{E}_k(\DK{r}_{1k})\,> \quad.
\EE
The first sum on the r.h.s of Eq.(\ref{div}) can be written in terms of the partial correlation function of particles $a$ and $b$, 
\be
<\ep^2>_1~=~g_{ab}(\DK{r})=\frac{1}{{V n_a n_b}} < \;\sum_{i=1}^{N_a} \sum_{j=1}^{N_b} \, \delta(\DK{r}\;-\;\DK{r}_{jb}\,+\,\DK{r}_{ia}) \;> \quad,
\ee
and reads
\be \label{diverg}
4 \pi n_e e^2 \int_0^{\infty} \, \frac{dr}{r^2} \; \left[ \, g_{ee}(r)\,+\,g_{ei}(r)\,\right] \quad,
\ee
which diverges at small distances, since for a fluid quantum system both $g_{ei}(0) \neq 0$ and $g_{ee}(0) \neq 0$. In the classical OCP only $g_{ee}(r)$ appears with $g_{ee}(0) = 0$, therefore the above integral and $<\ep^2>$ are finite. In contrast to what we have found Yan and Ichimaru \cite{YI86} predict a finite second moment. In Ref.\cite{YI86} no derivation of their   second moment expression valid ``strictly in the classical limit'' \cite{YI86} is given. To isolate a possible error in the derivation of Yan and Ichimaru one may perform semiclassical calculations of the second moment. Details of the semiclassical model are given in the next Section. We mention here only that in the semiclassical model the quantum system is modeled by a system of classical particles interacting via an effective potential $u_{ab}(r)=e_ae_b/r + u_{s,ab}(r)$, where the short-range part of the effective potential $u_{s,ab}(r)$ cuts the short-range divergency of the Coulomb potential. Therefore at short distances $u_{s,ab}(r \to 0)= -e_ae_b/r$. Within the semiclassical model the second moment reads
$$
<\ep^2>~=~ \frac{4 \pi n_e}{\beta} \left(g_{ei}(0)-g_{ee}(0) \right) - \beta \sum_{a,b} \left\langle \DK{\nabla}\frac{e_ae_b}{r}  \DK{\nabla} u_{s,ab}(r) \right\rangle \,.
$$
The first term in the above equation has been reported in Ref.\cite{YI86}, the second term has been omitted. However, as may be easily seen this second term is divergent. It may be expressed by an integral similar to that of Eq. (\ref{diverg}) which diverges at the lower bound. 

 Thus we have established a {\em qualitative} difference between the classical OCP and the TCP system. For the first the second moment of the microfield distribution is finite and corresponds to the variance of the Gaussian distribution. 

In contrast to the OCP case the second moment of the TCP system diverges. As a result the TCP microfield distribution does never converge to a Gaussian distribution.

We now generalize a coupling parameter technique which was used to calculate the microfield distribution of a classical OCP \cite{Iglesias,ILM83} to the case of a quantum TCP. Consider the function
\BEQ{Tl}
T(l)~&=&~\frac{Z(l)}{Z} \quad,\\
Z(l)~&\equiv&~{\rm Sp}\, e^{i \DK{l} \cdot \DK{E}} \,e^{-\beta \hat{H}} \quad.
\EEQ

Introduce the ``coupling strength'' parameter $\lambda$, $0 \leq \lambda \leq 1$, of the function, 
$$
Z(\lambda)= {\rm Sp}\, e^{i \lambda \DK{l} \cdot \DK{E}} \,e^{-\beta \hat{H}}\quad.
$$
From the definition of $T(l)$ in Eq. (\ref{Tl}) and assuming the first particle to be an electron one obtains
\BEQ{Tl1}
\ln T(l)~&=&~\int_0^1\, d\,\lambda\,\frac{\partial{ \ln Z(\lambda)}}{\partial \lambda} \nonumber \\
~&=&~\sum_a \, n_ae_a\, \int_0^1 \, d\,\lambda \,\int \, d \DK{r}\, \phi(\DK{r})\, g_{ea}(\DK{r},\lambda) \quad,
\EEQ
where
\BEQ{Twhere}
\phi(\DK{r})~&=&~\frac{i\,\DK{l} \cdot \DK{r}}{r^3} \quad , \\
g_{ab}(r,\lambda)~&=&~\frac{1}{{Z(\lambda) V n_a n_b}} {\rm Sp}   \;\sum_{i=1}^{N_a} \sum_{j=1}^{N_b} \, \delta(\DK{r}\;-\;\DK{r}_{jb}\,+\,\DK{r}_{ia})e^{i \lambda \DK{l} \cdot \DK{E}} \,e^{-\beta \hat{H}} \quad.
\EEQ
The functions $g_{ab}(r,\lambda)$ may be considered as generalized partial distribution functions.

In the case of a TCP Eq.(\ref{Tl1}) reads
\BE{TlTCP}
\ln T(l)~=~ \frac{n}{2} e \, \int_0^1 \, d\,\lambda \,\int \, d \DK{r}\, \phi(\DK{r})\, \left[ \, g_{ei}(\DK{r},\lambda)\, -\, g_{ee}(\DK{r},\lambda) \right] \quad.
\EE
The above expression is still exact. The use of the ``exponential approximation'' (EXP) \cite{ILM83} ansatz leads to the expression
\BE{APEX}
 g_{ea}(\DK{r},\lambda) ~\simeq ~g_{ea}(\DK{r},0) \, \exp\left[ {\cal E}_a(\DK{r};\lambda) \right] \,\,,~ a=e,i \quad,
\EE
with the ``renormalized potential'' given as \cite{ILM83}
\BEQ{APEXpot}
{\cal E}_a(\DK{r};\lambda)~&=&~{i \lambda \DK{l}} \cdot \DK{E}_a^*(\DK{r})\,,\nonumber \\
\DK{E}_a^*(\DK{r})&=&\DK{E}_a(\DK{r}_{1a})~+~\sum_b  n_b \int \, d\, \DK{r}_b \DK{E}_b(\DK{r}_{1b})\,\left[\,g_{ab}(r_{ab})-1\,\right] \quad.
\EEQ 
After substitution of Eq.(\ref{APEXpot}) into Eq.(\ref{APEX}) and performing the integration over $\lambda$ and the angles one gets
\ba
T(l)~&\simeq&~\exp\,\left[\,4 \pi \sum_a n_a \int_0^{\infty} \, dr\, r^2\, g_{ea}(\DK{r})\,\frac{E_a(r)}{E^*_a(r)}\,\left[j_0(lE^*_a(r))-1\right]\,\right] \quad, \label{EXP1} \\ 
E^*_a(r)~&=&~E_a(r)\,\left[\,1~+~4 \pi \sum_b  n_b \int_0^r \, r'^2 \, d\,{r}' \,\left[\,g_{ab}(r')-1\,\right] \right]\quad, \label{EXP2} \\
E_a(r)~&=&~\frac{e_a}{r^2} \label{EXP3}\quad,
\ea
with $j_0$ being the Bessel function of order zero. We notice that the use of the screened Coulomb potential Eq.(\ref{EXP2}) ensures the divergency of the second moment of the TCP microfield distribution. In this point our theory differs essentially from the results obtained by Yan and Ichimaru who used a potential of mean force instead of the screened Coulomb field \cite{YI86}. Eqs.(\ref{EXP1})-(\ref{EXP3}) constitute the so called exponential approximation (EXP) \cite {ILM83}. It is known that  in contrast to the so called adjustable parameter exponential approximation (APEX) the EXP expression poorly agrees with MD OCP data. In the APEX \cite {ILM83} one substitutes Eq. (\ref{EXP2}) by an {\em ad hoc} ansatz for $E^*_a(r)$. According to this ansatz the potential $E^*_a(r)$ is approximated by a parametrized Debye potential where the parameter is choosen to satisfy the second moment. In order to get a generalized APEX expression for the TCP one should know the second moment of the TCP microfield distribution. However, in the above consideration we have shown that the second moment of the TCP microfield distribution diverges. Therefore there is not any straightforward generalization of APEX to the TCP case. In the weak coupling limit both approximations, EXP and APEX, reduce to the Debye-H\"uckel (DH) approximation.

Consider therefore the DH limit in the case of TCP. In the weak coupling limit the pair correlation function is given by the screened pair approximation
 which in our case of a two-component plasma reads\cite{gruenes-buch}:
\BE{DHg}
g_{ab}(r)~=~S_{ab}^{(2)} (r)\,\exp\left[-\frac{\beta e_a e_b}{r} \left( e^{-\kappa r} - 1 \right) \right] \quad,
\EE
where $\kappa=(4 \pi \beta \sum_a n_a e_a^2)^{1/2}$ is the inverse Debye screening length. Further 
\BE{S2}
S_{ab}^{(2)} (r) = {\rm const.} {\sum_{\alpha}}' \exp\left( -\beta E_\alpha \right) 
   \mid \Psi_\alpha \mid^2 \;\;
\EE
 is the two-paricle Slater sum written in terms of the wave functions $\Psi_\alpha$ and energy levels $E_\alpha$ of the pair $ab$, respectively.  The prime at the summation sign indicates that the contribution of the bound states (which is not be considered here) has to be omitted. The Slater sum will be considered in the next section.

To calculate the effective field $E^*_a(r)$ in Eq.(\ref{EXP2}) it suffices to use the linear DH approximation
\BE{DHlinear}
g_{ab}(r)-1~=~-\frac{\beta e_a e_b}{r} \exp\left[-\kappa r\right] \quad,
\EE
since the nearest neighbour contribution to $E^*_a(r)$ is already singled out in Eq.(\ref{EXP2}). In addition, the linear DH approximation leads to a perfect screening of the impurity charge, which is an important requirement for a consistent approximation. The substitution of Eq.(\ref{DHlinear}) into Eq.(\ref{EXP2}) yields the Debye screened field
\BE{DHfield}
E^*_a(r)~=~\frac{e_a}{r^2}\,\left(1+\kappa r \right) \, \exp(-\kappa r) \quad.
\EE
We put now Eqs. (\ref{DHfield}) and (\ref{DHg}) into Eq. (\ref{EXP1}) and obtain the DH approximation for the microfield distribution in a two-component plasma. This approximation may be expressed in terms of the dimensionless quantities introduced in Eqs. (\ref{dimless1}) and (\ref{PF}) and reads
\BEQ{DHA}
T(L)~&=&~T_{ee}(L)\,T_{ei}(L) \quad, \nonumber \\
\ln \, T_{ea}(L)~&=&~ \frac{15 L^{3/2}}{4 \sqrt{2\pi}}\;\int_0^{\infty}\frac{dx}{x^2 B(x)}\;\left(\, \frac{\sin \, B(x)}{B(x)}\;-\;1\,\right)\,\exp\left[\,Z_a\frac{\Gamma c}{\sqrt{L}}\,e^{-\sqrt{6 \Gamma L}/cx} \, \right] \nonumber \\
&&\;~\cdot ~\exp\left(\beta\,u_{s,ea}(\frac{d \sqrt{L}}{cx}) \right)\quad, \nonumber\\
B(x)\;&=&\;x^2\;\left(1\;+\;\frac{\sqrt{6 \Gamma L}}{cx}\,\right)\;\exp\left[-\frac{\Gamma c}{\sqrt{L}}\,\right]\;,~c=\frac{\sqrt{2\pi}\,2^{1/3}}{(5 \pi )^{1/3}}\,,~Z_i=-Z_e=1 \quad,
\EEQ
where the electron Wigner-Seitz radius $d$ and the coupling constant $\Gamma$ have been defined in the Introduction. Further in Eq. (\ref{DHA}) we have introduced an effective short range potential
$$
\exp \left(- \beta\,u_{s,ea}(r) \right) = S_{ea}^{(2)} (r)\, \exp\left(-\frac{\beta e e_a}{r} \right) \quad.
$$
Equation (\ref{PF}) with $T(L)$ from Eq. (\ref{DHA}) constitutes the Debye-H\"uckel approximation for the microfield distribution applicable to the weakly coupled TCP. These equations generalize the corresponding DH approximation used to calculate the OCP microfield distribution \cite{Iglesias}. We mention that the approximation Eqs. (\ref{DHA}) can be directly obtained from Eq. (\ref{TlTCP}) using the nonlinear Debye-H\"uckel approximation for the generalized radial distribution function,
\BE{DHglambda}
g_{ea}(r;\lambda)~=~\exp\left[\, \beta \left[ 1+\frac{i \lambda l \DK{\nabla}}{e \beta}\, \right] \frac{e e_a}{r} e^{-\kappa r}\right]\,\,\exp\left[\,-\beta u_{s,ea}(r)\right] \quad.
\EE

In the next section we consider the two-particle Slater sum and introduce the semiclassical model employed in the numerical simulations.

\section{Slater sum, semiclassical model and MD-Simulations} \label{model}

As pointed out in the Introduction the idea of the semiclassical methods is to incorporate
quantum-mechanical effects (in particular the
Heisenberg and the Pauli principle) by appropriate effective potentials.

An easy way to arrive at effective potentials describing quantum effects is the use of
the so-called Slater sums which 
were studied in detail by several authors \cite{gruenes-buch,ebeling}. 
The Slater sum caracterizes the distribution
of the system in coordinate space. Choosing the logarithm of the Slater sum

\be
U^{(N)}(\DK{r}_1,\ldots,\DK{r}_N ) = - T \ln S( \DK{r}_1,\ldots,\DK{r}_N ) \quad ,
\label{slater}
\ee
as a potential for the classical motion of the particles, we map our quantum
system onto a classical one.
The potentials $U^{(N)}(\DK{r}_1,\ldots,\DK{r}_N )$ are often called quantum statistical effective potentials
and they are used to calculate the correct thermodynamic functions of the original quantum system \cite{gruenes-buch,ebeling,norman}.

The Slater sum may be considered as an analogue of the classical Boltzmann factor. 
The only modification in comparison with the classical theory is the appearance of many-particle interactions. If the system is not to dense (i.e., in the nondegenerate limit) one may neglect the contributions of higher order many-particle interactions. In this case one writes approximately,

\BE{pair}
U^{(N)}(\DK{r}_1,\ldots,\DK{r}_N ) \approx \sum_{i<j} u_{ij}(\DK{r}_i,\DK{r}_j) \quad,
\EE
where the effective two-particle potential $u_{ab}$ is defined by the two-particle Slater sum Eq. (\ref{S2}).

The Slater sum for the pair of charged particles can be approximated in
different ways. Following Kelbg \cite{Kelbg} one considers the Coulombic interaction as a perturbation; in the first order one gets the expression
\begin{equation}\label{Kelbg-pot}
u_{ab}(r)={e_ae_b\over{r}}\Big(F (r / \lambda_{ab})\Big),
\end{equation}
with
\begin{equation}\label{Fx}
F(x) = 1- \exp\left(-x^2\right) +\sqrt{\pi}  x \left(1-
\mbox{erf}\left( x \right) \right) \quad,
\end{equation}
which we will call the Kelbg potential. Here $\lambda_{ab}=\hbar/\sqrt{2 m_{ab}T}$ is De Broglie wave length of relative motion, $m_{ab}^{-1}=m_a^{-1}+m_{b}^{-1}$, $a=e,i$, $m_e$ and $m_i$ being the electron and ion masses, respectively. Further in Eq.(\ref{Kelbg-pot}) we have neglected the exchange contributions. An effective potential similar to Eq. (\ref{Kelbg-pot}) was 
derived by Deutsch and was used in the simulations by Hansen and McDonald 
\cite{hansen2}. 

The Kelbg potential is a good approximation for the two-particle Slater sum in the case of small parameters $\xi_{ab} = - (e_a e_b)/(T \lambda_{ab})$ if the interparticle distance $r$ is sufficiently large.  At small interparticle distances it deviates from the exact value of $- T \cdot \ln ( S_{ab}(r=0))$. In order to describe the right behavior at small distances it is better to use a corrected Kelbg potential defined by \cite{VE99,JVE99}

\begin{equation} \label{corr-Kelbg}
u_{ab}(r) =
\left({e_a e_b}/{r}\right)\cdot  F (r / \lambda_{ab})  - k_B T 
\tilde{A}_{ab}(\xi_{ab}) \exp
\left(-(r / \lambda_{ab})^2 \right)  \quad.
\end{equation}

In Eq. (\ref{corr-Kelbg}) the temperature-dependent magnitude
$\tilde{A}_{ab}(T)$
is adapted in such a way that the Slater sum ${S_{ab}(r=0)}$ and its first derivative ${S'_{ab}(r=0)}$ have the exact value at zero distance known from previous works 
 \cite{gruenes-buch,Rohde}. The explicit expressions read \cite{JVE99}
\BEQ{Aee}
\tilde{A}_{ee}&=&\sqrt{\pi} |\xi_{ee}| +\ln \left[2 \sqrt{\pi} |\xi_{ee}| \, \int \frac{dy\,y \exp \left(-y^2 \right)}{\exp\left( \pi |\xi_{ee}|/y \right)-1} \right]
\\
\tilde{A}_{ei}&=&- \sqrt{\pi}\xi_{ei}+\ln \left[ \sqrt{\pi} 
\xi_{ie}^3 
\left( \zeta(3) + \frac{1}{4} \zeta(5) \xi_{ie}^2 \right) \right.
+ \left. 4 \sqrt{\pi} \xi_{ei} \, \int \frac{dy\,y \exp \left(-y^2 \right)}{1-\exp{\left(-\pi \xi_{ei}/y \right)}} \right]
\EEQ

For low temperatures $0.1 < T_r < 0.3$
one shall  use the corrected Kelbg-potential Eq.(\ref{corr-Kelbg}) to get an appropriate approximation for the Slater sum at arbitrary distances. 

In the region of higher temperatures
\begin{equation}
T_r = T/T_I = \left({2 T \hbar^2}/{m_{ie} e^4}\right) > 0.3 \quad
\end{equation}
the Kelbg potential ($A_{ab}=0$) and the corrected Kelbg potential almost coincide. At still higher temperatures $T/T_I>1$ the Kelbg potential does not differ from the
corrected Kelbg potential only in the case of
electron-ion interaction.
For the interaction of the particles of the same type the correction $\tilde{A}_{ab}$
includes also the exchange effects
, which
make the potential
unsymmetrical (that means $u_{ei}$ differ from  $u_{ee}$). The 
potential assymetry becomes
apparent at  high temperatures (${{T}} > 100000\; {\rm{K}}$) only.

In the present work
we are interested in the regime of intermediate temperatures. Therefore the simulations are performed with the potential Eq.(\ref{corr-Kelbg}) which is presented in Fig. 1 and compared with other potentials approximating the two-particle Slater sum. 

To check the quality of the predictions from the approximation given in Sec. \ref{II} we have performed a series of molecular dynamic simulations for comparison. The leap-frog
variant of Verlet's algorithm was used to integrate numerically the equations of motions obtained from the effective potential Eq.(\ref{corr-Kelbg}). 
The simulations were performed using a 256-particle system of electrons and positrons with periodical
boundary conditions. The temperature of the system was choosen as $T=30\,000~{\rm K}$, the coupling has varied from weak coupling ($\Gamma =0.2$) up to intermediate coupling ($\Gamma =2$). In the investigated range of plasma parameters the size of the simulation box was significantly greater than the Debye radius. Therefore the long-range Coulomb forces are screened inside each box and no special 
procedure like Ewald summation was implemented to calculate them.  Either 
MD runs with Langevin source or MC procedures were used to establish thermal
equilibrium in the system, both methods have led to the same results. 

In Figs. 2-5 we present the results of the approximation Eqs. (\ref{PF}) and (\ref{DHA}) as well as the Holtsmark (Eq. (\ref{Holts})) approximation. The short range potential in Eq. (\ref{DHA}) is given by the corrected Kelbg potential without the Coulomb term
\begin{equation} \label{short-pot}
u_{s,ab}(r) =
\left({e_a e_b}/{r}\right)\cdot \left\{(F (r/\lambda_{ab})-1)\right\} +
A_{ab} \exp
(-(r/\lambda_{ab})^2)  \quad,
\end{equation}
with $F(x)$ from Eq.(\ref{Fx}).

The results of the analytical approximation are compared with MD data. It can be seen from the figures that the Debye-H\"uckel approximation is in good agreement with the MD data for the case of weak coupling, however, with increasing coupling strength this agreement becomes poorer. This is not surprising, since the DH approximation is valid only in the weak coupling regime. To get a better agreement for the case of intermediate coupling one has to improve the calculation of the radial distribution function. 

From the figures we also see that the MD data show a large probability of high microfield values. The long tails in the distribution function reflect the attraction between oppositely charged particles. As a result the probability to find a particle of opposite charge at small distances from the probe charge and thus producing large microfields is even higher than in the ideal Holtsmark case. This situation is in striking contrast to the OCP case where the repulsion of particles with the same charge leads to a small probability of high microfield values. As for the TCP the long tails are still present in the case of an intermediate coupling for which the OCP microfield distribution approaches the Gaussian distribution Eq. (\ref{Gauss}) \cite{ILM83}. In the DH approximation the long tails are less pronounced for the case $\Gamma=2$. Here the Debye-H\"uckel length is smaller than the average distance between the particles. Thus the particle interactions become screened even at short distances. A  result of this unphysical screening is the supression of high microfields within the DH approximation and for large coupling parameters. At still higher densities ($\Gamma \ge 3$ at $T=30\,000~{\rm K}$) the De-Broglie wavelength becomes comparable with the interparticle distance and the semi-classical approach employed in the numerical part of the paper fails to describe the quantum two-component plasma properly.

\section{Conclusions}

The electric microfield distribution at a charged particle in a two-component plasma has been studied. Generalizing the corresponding transformation for the case of a classical OCP we have expressed the Fourier transform of the electric microfield distribution in terms of generalized partial radial distribution functions. Using a simple Debye-H\"uckel like generalized radial distribution function (including the unscreened short range part stemming from the effective potential) we have obtained theoretical predictions for the electric microfield distribution of the TCP. It has been shown that in contrast to the OCP the second moment of the TCP microfield distribution diverges. 

Semiclassical molecular-dynamics simulations of the two-component plasma using effective potentials have been performed . The effective potential was choosen to describe the nondegenrate limit of the quantum system appropriately. The microfield distribution for different coupling constants (from $\Gamma=0.2$ to $\Gamma=2.0$) has been obtained. With increasing coupling strength the most probable value of electric microfields is shifted to lower fields. However, at all coupling strengths for which the simulations have been performed the microfield distribution shows long tails indicating a large probability of high microfields. This behavior is in contrast to the corresponding behavior in one-component plasmas. It reflects the divergency of the second moment of the TCP microfield distribution.

 At weak coupling there is an overall agreement of the microfield distribution obtained by the analytical approximation with the MD data. Although our simple approximation fails to provide accurate numerical results for larger coupling constants, the formalism allows to generalize the results to the case of intermediate and strong coupling.

\section{Acknowledgments}

This work was supported by the Deutsche Forschungsgemeischaft (DFG, Germany) and the Deutscher Akademischer Austauschdienst (DAAD, Germany).


\newpage

\begin{center}
{\bf FIGURE CAPTIONS}
\end{center}

\begin{description}

\item[(Figure 1)] Effective potentials Eq.(\ref{Kelbg-pot})(Kelbg potential) and Eq.(\ref{corr-Kelbg}) (corrected Kelbg potential). The Kelbg potential is drawn for three temperatures, the corrected Kelbg-potential is explicitely shown at $T=10\,000~{\rm K}$ for both interactions and at $T=100\,000~{\rm K}$ for the electron-electron interaction only;  in the other cases the corrected Kelbg potential coincides with the Kelbg potential within the figure accuracy. For comparison we have included also the low-temperature limit of the effective potential of free charges (the ``classical'' potential-dashed line); the repulsive part of the classical potential coincides with the bare Coulomb potential.
\item[(Figure 2)] Comparison of microfield distribution $P(F)$ curves at $T = 30\;000 {\rm K}$ and $\Gamma=0.2$ from molecular dynamics (MD) and the analytical approximation derived in this work (DH) Eqs. (\ref{PF}) and (\ref{DHA}).
\item[{Figure 3}] Same as in Fig. 2 at $\Gamma=0.8$.
\item[{Figure 4}] Same as in Fig. 2 at $\Gamma=1.2$.
\item[{Figure 5}] Same as in Fig. 2 at $\Gamma=2.0$.
\end{description}


\vspace*{-2cm}
\begin {figure} [h] 
\unitlength1mm
  \begin{picture}(145,150)
\put (0,-10){\psfig{figure=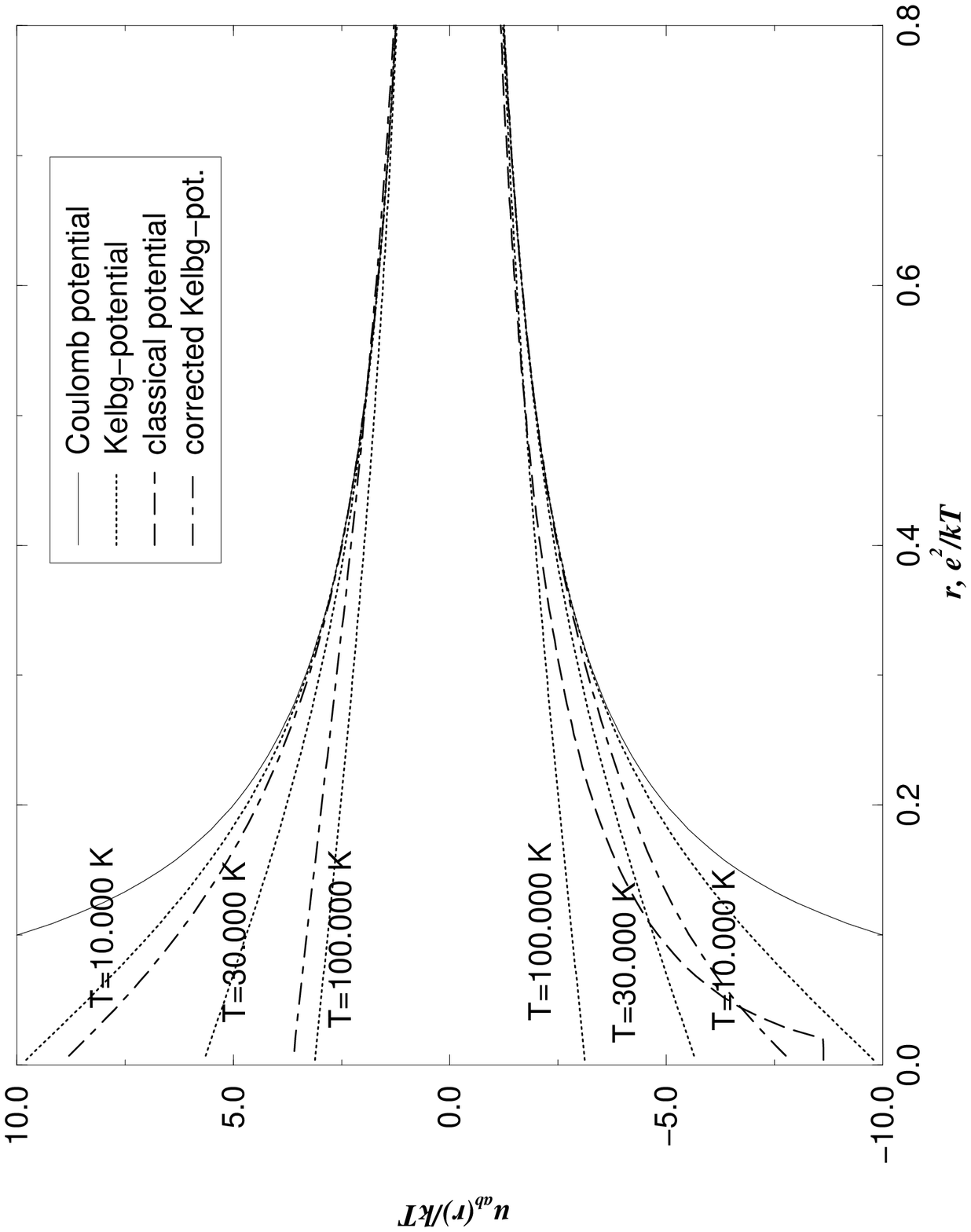,width=14.5cm,height=14.0cm,angle=0}}
 \end{picture}\par
\end{figure}

\vspace*{4.5cm}
Figure 1. (Microfield distribution in two-component plasmas; Ortner, Valuev, Ebeling)

\newpage

\vspace*{-2cm}
\begin {figure} [h] 
\unitlength1mm
  \begin{picture}(145,160)
\put (0,-10){\psfig{figure=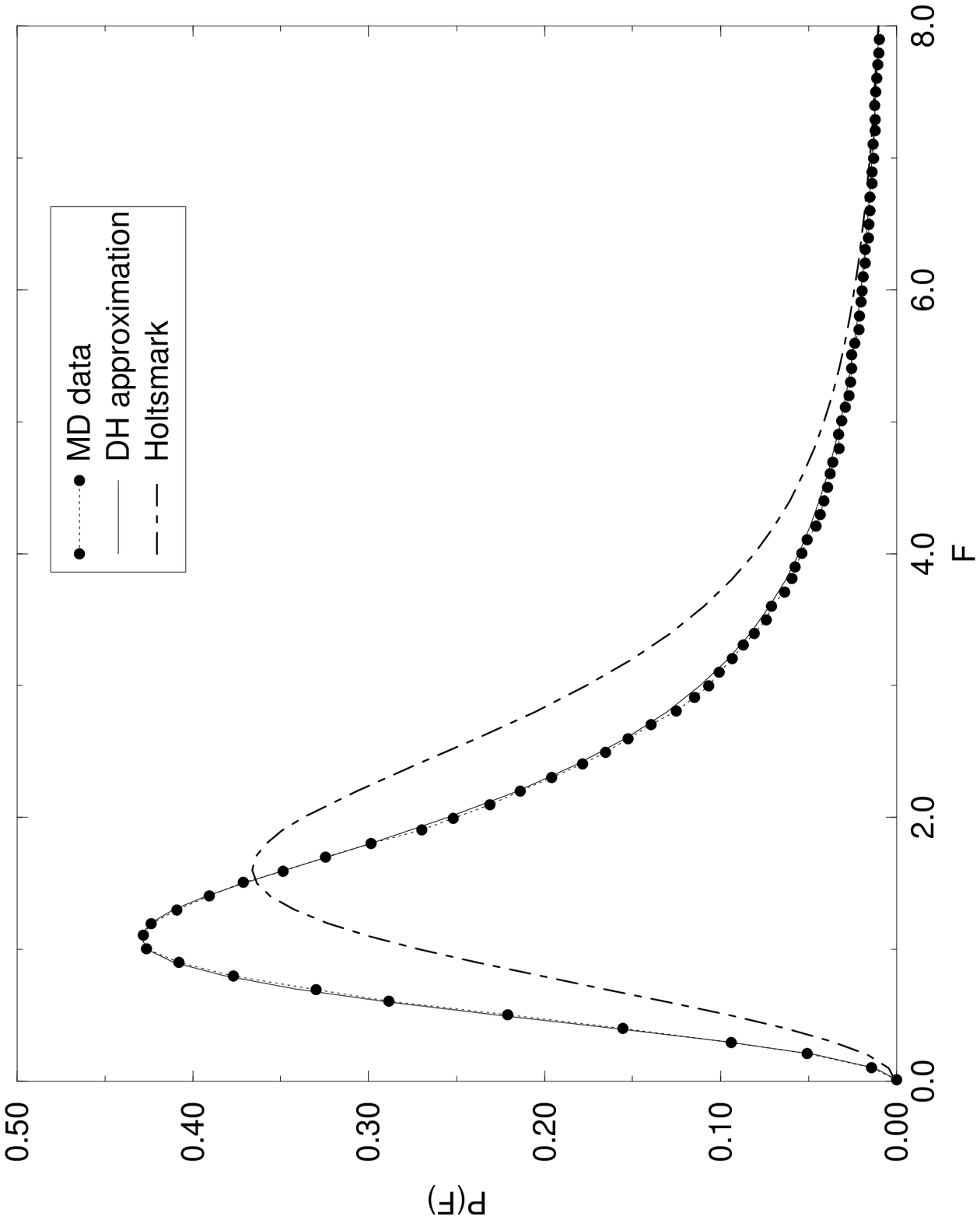,width=14.5cm,height=14.0cm,angle=0}}
 \end{picture}\par 
\end{figure}

\vspace*{4.5cm}
Figure 2.  (Microfield distribution in two-component plasmas; Ortner, Valuev, Ebeling)

\newpage

\vspace*{-2cm}
\begin {figure} [h] 
\unitlength1mm
  \begin{picture}(145,160)
\put (0,-10){\psfig{figure=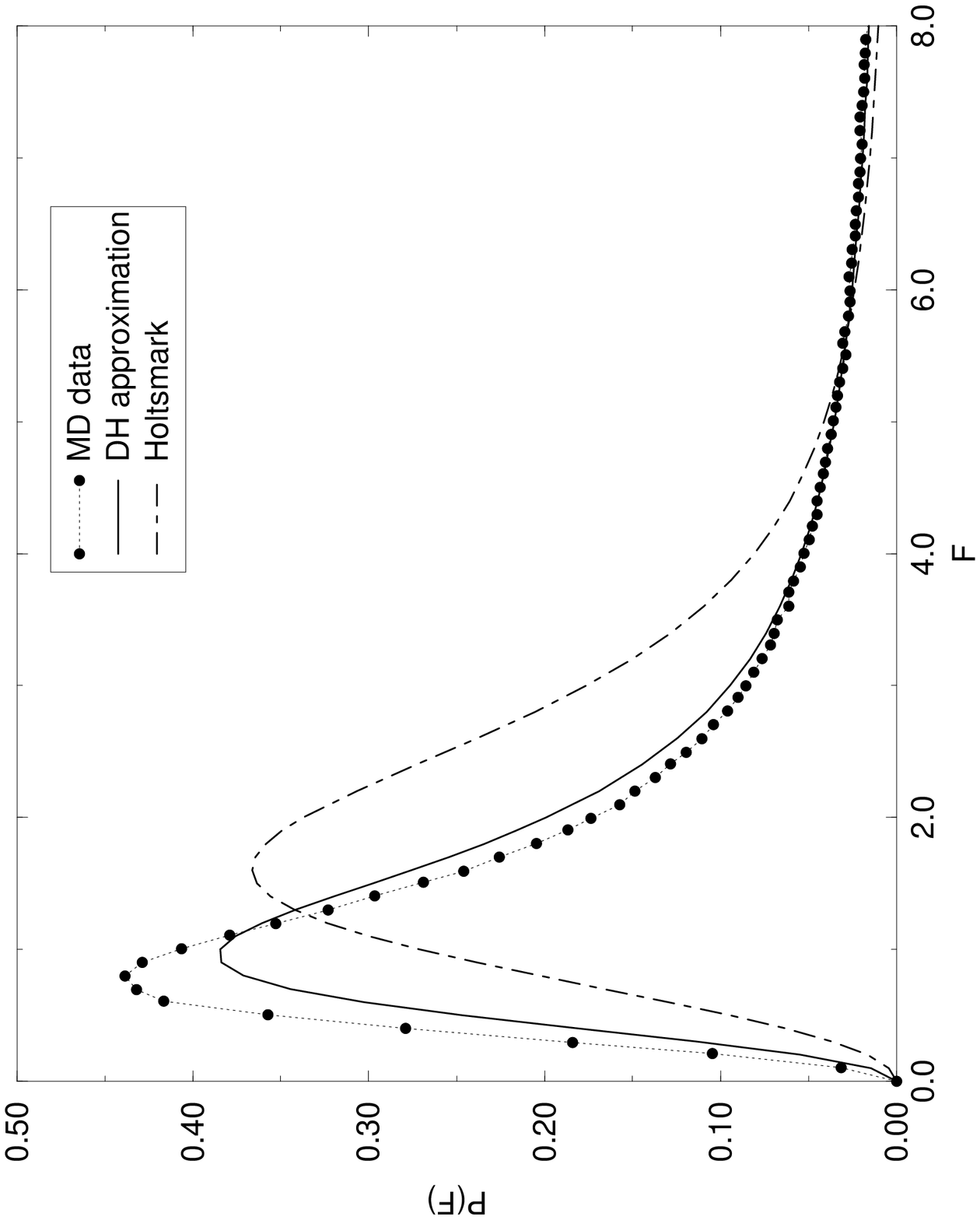,width=14.5cm,height=14.0cm,angle=0}}
 \end{picture}\par 
\end{figure}

\vspace*{4.5cm}
Figure 3.  (Microfield distribution in two-component plasmas; Ortner, Valuev, Ebeling)

\newpage

\vspace*{-2cm}
\begin {figure} [h] 
\unitlength1mm
  \begin{picture}(145,160)
\put (0,-10){\psfig{figure=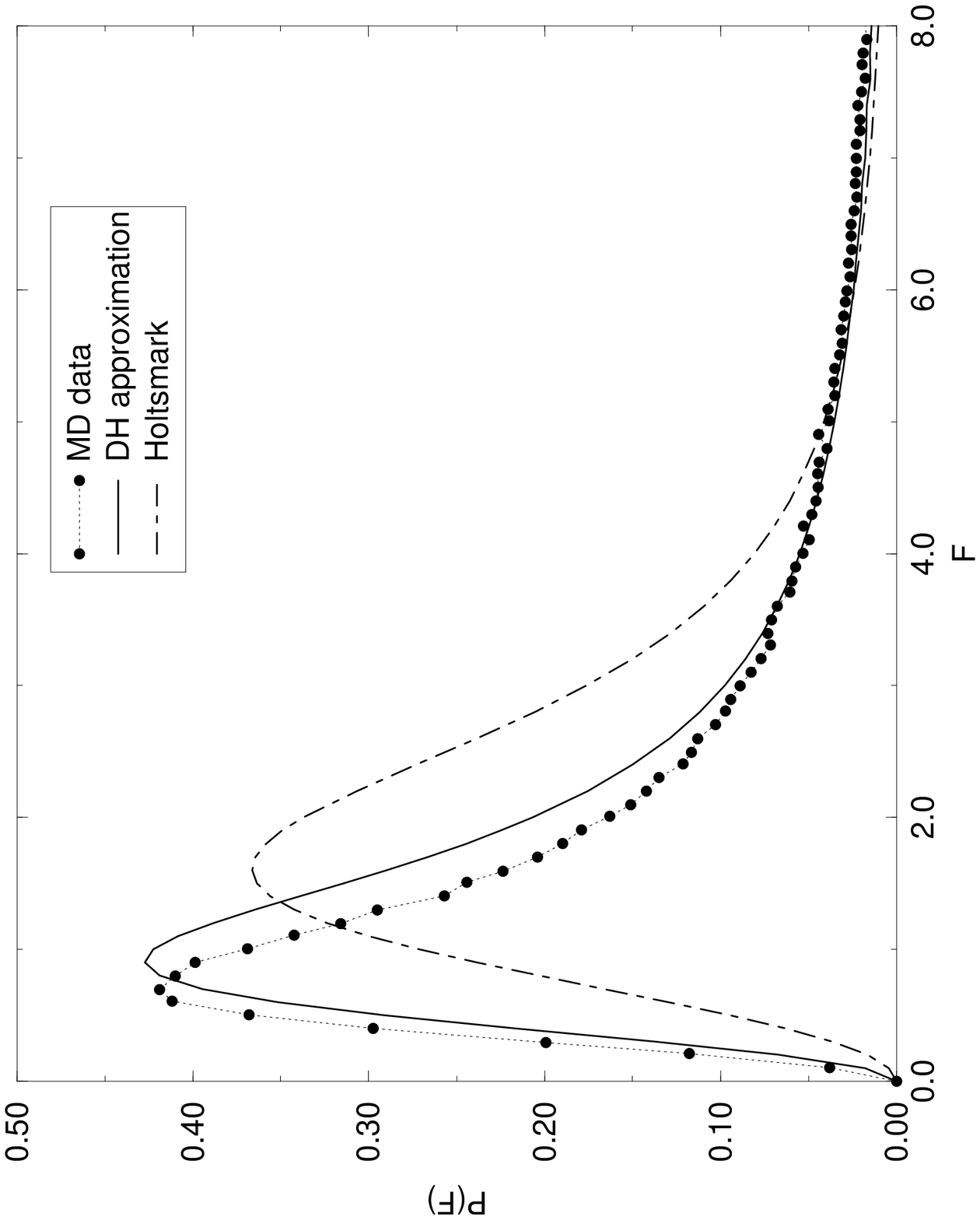,width=14.5cm,height=14.0cm,angle=0}}
 \end{picture}\par 
\end{figure}

\vspace*{4.5cm}
Figure 4.  (Microfield distribution in two-component plasmas; Ortner, Valuev, Ebeling)

\newpage

\vspace*{-2cm}
\begin {figure} [h] 
\unitlength1mm
  \begin{picture}(145,160)
\put (0,-10){\psfig{figure=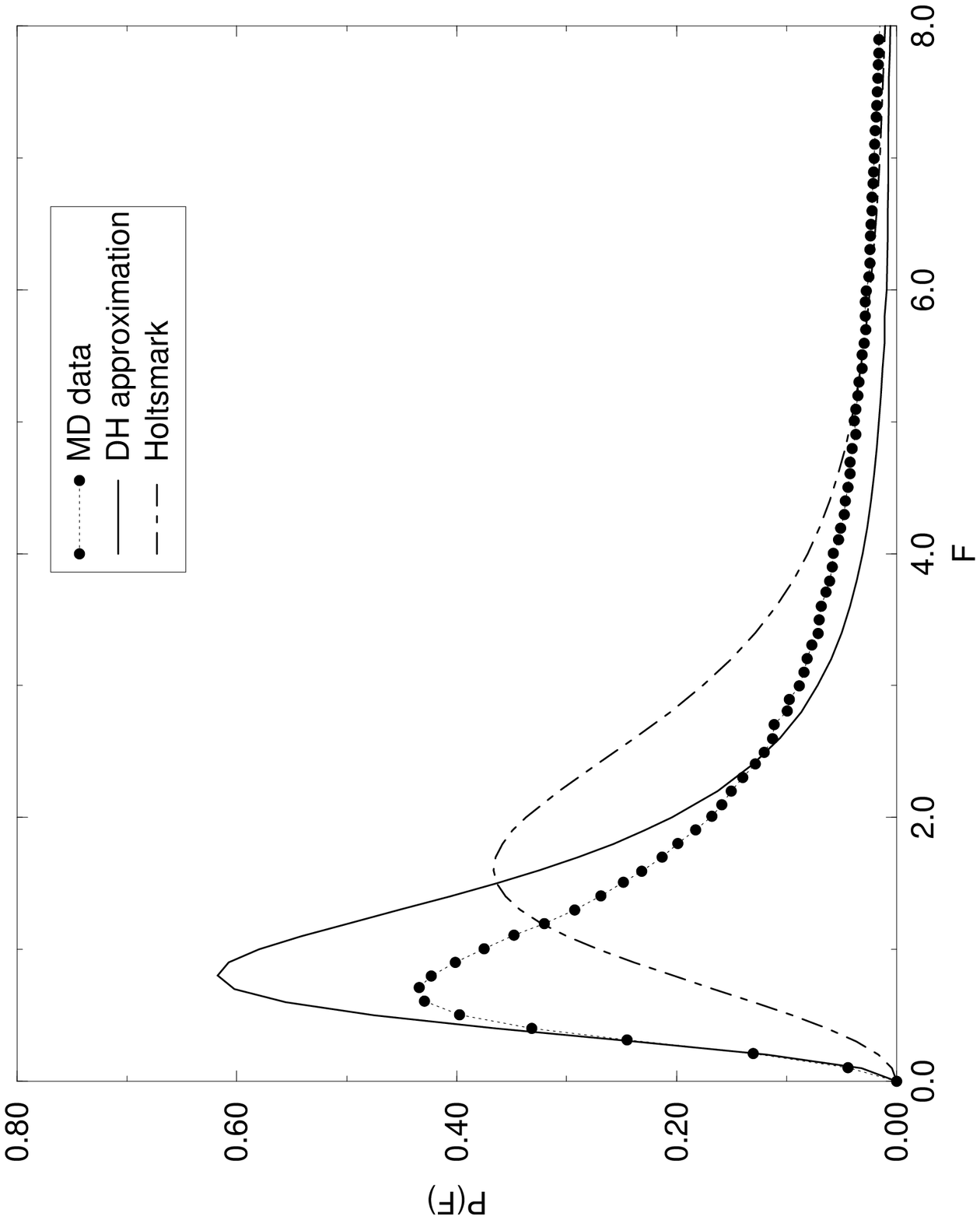,width=14.5cm,height=14.0cm,angle=0}}
 \end{picture}\par 
\end{figure}

\vspace*{4.5cm}
Figure 5.  (Microfield distribution in two-component plasmas; Ortner, Valuev, Ebeling)

\end{document}